\documentclass[a4paper,10pt,onecolumn]{article}

\usepackage{mathptmx}
\usepackage{url}
\usepackage{afterpage}
\usepackage{graphicx}
\usepackage{epsfig}
\usepackage{setspace}

\begin{document}


\title{Collective Mind: cleaning up the research and experimentation mess in computer engineering using crowdsourcing, big data and machine learning}

\author{
Grigori Fursin  \\
{\itshape INRIA, France}\\
\small{Grigori.Fursin@cTuning.org}
}

\date{}

\maketitle

\begin{abstract}

Software and hardware co-design and optimization of HPC systems has become
intolerably complex, ad-hoc, time consuming and error prone due to enormous
number of available design and optimization choices, complex interactions
between all software and hardware components, and multiple strict
requirements placed on performance, power consumption, size, reliability
and cost.

We present our novel long-term holistic and practical solution to this
problem based on customizable, plugin-based, schema-free, heterogeneous,
open-source Collective Mind repository and infrastructure with unified web
interfaces and on-line advise system. This collaborative framework distributes 
analysis and multi-objective off-line and on-line auto-tuning of computer systems among many participants
while utilizing any available smart phone, tablet, laptop, cluster or data
center, and continuously observing, classifying and modeling their
realistic behavior. Any unexpected behavior is analyzed using shared data
mining and predictive modeling plugins or exposed to the community at
cTuning.org for collaborative explanation, top-down complexity reduction,
incremental problem decomposition and detection of correlating program,
architecture or run-time properties (features). Gradually increasing
optimization knowledge helps to continuously improve optimization
heuristics of any compiler, predict optimizations for new programs or
suggest efficient run-time (online) tuning and adaptation strategies depending on end-user
requirements. We decided to share all our past research artifacts including 
hundreds of codelets, numerical applications, data sets, models, universal 
experimental analysis and auto-tuning pipelines, self-tuning machine learning 
based meta compiler, and unified statistical analysis and machine learning plugins
in a public repository to initiate systematic,
reproducible and collaborative research, development and experimentation with a new publication model where
experiments and techniques are validated, ranked and improved by the community.

\end{abstract}

\textbf{Keywords}
\small{
\textit{
Collective Mind,  crowdtuning,  crowdsourcing auto-tuning and co-design,  software and hardware co-design and co-optimization,
on-line tuning and learning,  systematic behavior modeling,  predictive modeling,  data mining,  machine learning, 
on-line advice system,  metadata,
top-down optimization,  incremental problem decomposition,  decremental (differential) analysis,
complexity reduction,  tuning dimension reduction,  customizable plugin-based infrastructure,
public repository of knowledge,  big data processing and compaction,  agile research and development, 
cTuning.org,  c-mind.org ,  systematic and reproducible research and experimentation,  validation by community}
}




\section{Introduction, major challenges, and related work}

\subsection{Motivation}

Continuing innovation in science and technology is vital for our society
and requires ever increasing computational resources. However, delivering
such resources particularly with exascale performance for HPC or ultra low power
for embedded systems is becoming intolerably complex, costly and error prone due
to limitations of available technology, enormous number of available design and 
optimization choices, complex interactions between all software and hardware components,
and growing number of incompatible tools and techniques with ad-hoc,
intuition based heuristics. As a result, understanding and modeling of the overall relationship between
end-user algorithms, applications, compiler optimizations, hardware
designs, data sets and run-time behavior, essential for providing better
solutions and computational resources, became simply infeasible as
confirmed by numerous recent long-term international research visions about
future computer systems~\cite{citeulike:1671417,hipeac_roadmap,Dongarra:2011:IES:1943326.1943339,DBLP:books/ms/4paradigm09,uhpc,prace}.
On the other hand, the research and development methodology for computer
systems has hardly changed in the past decades: computer architecture is
first designed and later compiler is being tuned and adapted to the new
architecture using some ad-hoc benchmarks and heuristics. As a result, peak
performance of the new systems is often achieved only for a few previously
optimized and not necessarily representative benchmarks such as SPEC
for desktops and servers or LINPACK for TOP500 supercomputer ranking,
while leaving most of the systems severely underperforming and wasting
expensive resources and power. 

Automatic off-line and on-line performance tuning techniques were introduced nearly two decades
ago in an attempt to solve some of the above problems. These approaches treat 
computer systems as a black box and explore their optimization parameter space
empirically, i.e. compiling and executing a user program multiple times
with varying optimizations or designs (compiler flags and passes, fine-grain
transformations, frequency adaptation, cache reconfiguration, parallelization, etc) 
to empirically find better solutions that improve execution and compilation
time, code size, power consumption and other characteristics~\cite{atlas,fftw,pfdc,CSS99,KKO2000,FOK02,CST02,vista,TVVA03,spiral,PE2004,PE2006,HE2008}.
Such techniques require little or no knowledge of the current platform and
can adapt programs to any given architecture automatically. With time,
auto-tuning has been accelerated with various adaptive exploration techniques
including genetic algorithms, hill-climbing and probabilistic focused 
search. However, the main disadvantage of these techniques is an excessively long
exploration time of large optimization spaces and lack of optimization
knowledge reuse among different programs, data sets and architectures.
Moreover, all these exploration steps (compilation and execution) must be
performed with exactly the same setup by a given user including the same
program, generated with the same compiler on the same architecture with the
same data set, and repeated a large number of times to become statistically
meaningful.

Statistical analysis and machine learning have been introduced nearly a decade ago to
speed up exploration and predict program and architecture behavior,
optimizations or system configurations by automatically learning
correlations between properties of multiple programs, data sets and
architectures, available optimizations or design choices, and
observed characteristics~\cite{atlas,fftw,pfdc,CSS99,KKO2000,FOK02,CST02,vista,TVVA03,spiral,PE2004,PE2006,HE2008,DBLP:journals/taco/TartaraC13,29db2248aba45e59:b254c18c8794ba29,29db2248aba45e59:530e5f456ea259de}.
Often used by non-specialists, these approaches mainly demonstrate a potential 
to predict optimizations or adaptation scenario in some limited cases, but
they do not include deep analysis about machine learning algorithms, their 
selection and scalability for ever growing training sets, optimization choices and
available features which are often problem dependent, are the major
research challenges in the field of machine learning for several decades,
and far from being solved.

We believe that many of the above challenges and pitfalls are caused
by the lack of a common experimental methodology, lack of interdisciplinary
background, and lack of unified mechanisms for knowledge building and exchange 
apart from numerous similar publications where 
reproducibility and statistical  meaningfulness of results as well as sharing of data 
and tools is often not even considered in contrast with other sciences including 
physics, biology and artificial intelligence. In fact, it is often impossible due to a lack of common 
and unified repositories, tools and data sets. At the same time, there is a vicious
circle since initiatives to develop common tools and repositories to unify,
systematize, share knowledge (data sets, tools, benchmarks, statistics,
models) and make it widely available to the research and teaching community
are practically not funded or rewarded academically where a number of
publications often matter more than the reproducibility and statistical
quality of the research results. As a consequence, students, scientists and engineers are forced to resort to some
intuitive, non-systematic, non-rigorous and error-prone techniques combined
with unnecessary repetition of multiple experiments using ad-hoc tools,
benchmarks and data sets. Furthermore, we witness slowed down innovation,
dramatic increase in development costs and time-to-market for the new
embedded and HPC systems, enormous waste of expensive computing resources
and energy, and diminishing attractiveness of computer engineering often
seen as "hacking" rather than systematic science.

\section{Collective Mind approach}

\subsection{Back to basics}

We would like to start with the formalization of the eventual needs of
end-users and system developers or providers. End-users generally need to perform
some tasks (playing games on a console, watching videos on mobile or
tablet, surfing Web, modeling a new critical vaccine on a supercomputer or
predicting a new crash of financial markets using cloud services) either as
fast as possible or with some real-time constraints while minimizing or
amortizing all associated costs including power consumption, soft and hard
errors, and device or service price. Therefore, end-users or adaptive
software require a function that can suggest most optimal design or
optimization choices~\textbf{c} based on properties of their tasks and data
sets~\textbf{p}, set of requirements~\textbf{r}, as well as current state
of a used computing system~\textbf{s}:

\begin{displaymath} 
\bf{c} = \emph{F}(\bf{p},\bf{r},\bf{s})
\end{displaymath}

This function is associated with another one representing behavior of a user
task running on a given system depending on properties and choices:

\begin{displaymath} 
\bf{b} = \emph{B}(\bf{p},\bf{c},\bf{s})
\end{displaymath}

This function is of particular importance for hardware and software
designers that need to continuously provide and improve choices (solutions)
for a broad range of user tasks, data sets and requirements while trying to
improve own ROI and reduce time to market. In order to find optimal
choices, it should be minimized in presence of possible end-user
requirements (constraints). However, the fundamental problem is that
nowadays this function is highly non-linear with  such a multi-dimensional
discrete and continuous parameter space which is not anymore possible to
model analytically or evaluate empirically using exhaustive search~\cite{europar97,atlas,29db2248aba45e59:61df8131994fad97}.
For example, \textbf{b} is a behavior vector that can now include execution
time, power consumption, compilation time, code size, device cost, and any
other important characteristic; \textbf{p} is a vector of properties of a
task and a system that can include semantic program features~\cite{Monsifrot,SAMP2003,ABCP06,29db2248aba45e59:a31e374796869125},
dataset properties, hardware counters~\cite{CFAP2007,JGVP2009}, system configuration, and
run-time environment parameters among many others; \textbf{c} represents
available design and optimization choices including algorithm selection,
compiler and its optimizations, number of threads, scheduling, processor
ISA, cache sizes, memory and interconnect bandwidth, frequency, etc; and finally
\textbf{s} represents the state of the system during parallel execution of
other programs, system or core frequency, cache contentions and so on.

\subsection{Interdisciplinary collaborative methodology}

Current multiple research projects mainly show that it is possible to use
some off-the-shelf on-line or off-line adaptive exploration (sampling)
algorithms combined with some existing models to approximate above function
and predict behavior, design and optimization choices for 70-90\% cases but
in a very limited experimental setup. In contrast, our ambitious long-term
goal is to understand how to continuously build, enhance, systematize and
optimize hybrid models that can \emph{explain and predict all possible
behaviors and choices} while selecting minimal set of representative
properties, benchmarks and data sets for predictive modeling~\cite{LCWP2009}.
We reuse our interdisciplinary knowledge in physics, quantum electronics 
and machine learning to build a new methodology that can
effectively deal with rising complexity of computer systems through 
gradual and continuous top-down problem decomposition, analysis and learning.
We also develop a modular infrastructure and repository that allows
to easily interconnect various available tools and techniques to
distribute adaptive probabilistic exploration, analysis and optimization of computer
systems among many users~\cite{software-cm,cm-repo} while exposing unexpected or unexplained behavior
to the community with interdisciplinary backgrounds particularly in
machine learning and data mining through unified web interfaces 
for collaborative solving and systematization.

\subsection{Collective Mind infrastructure and repository}

\begin{figure}[ht]
  \centering
  \includegraphics[width=4.4in, bb=40 18 668 504]{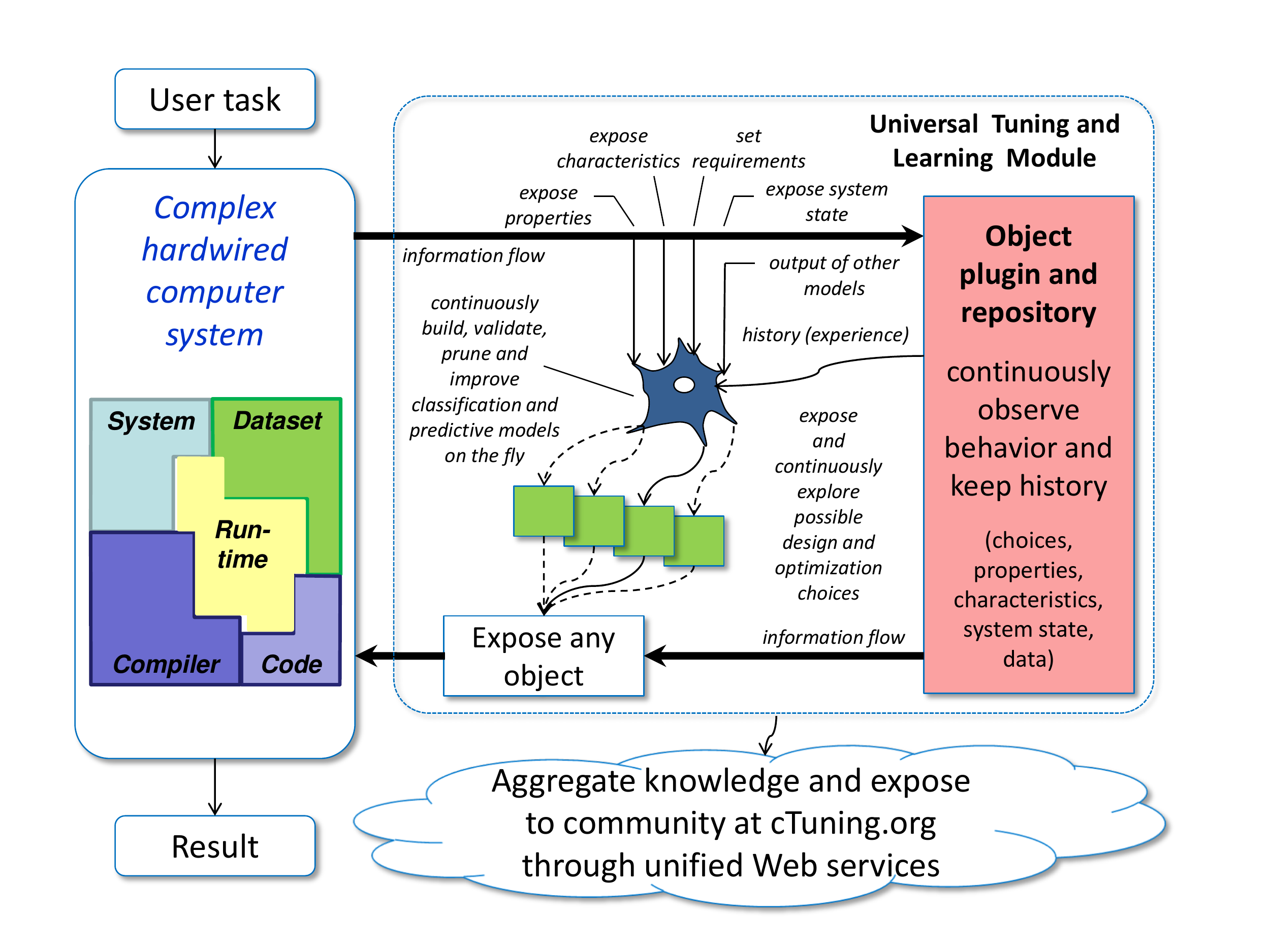}
  \caption{\it Gradual decomposition, parameterization, observation, tuning and learning of complex hardwired computer systems.}
  \label{fig:universal-node}
\end{figure}

Collective Mind framework and repository (cM for short) enables continuous, 
collaborative and agile top-down decomposition of the whole complex hardwired 
computer systems into unified and connected subcomponents (modules) with 
gradually exposed various characteristics, tuning choices (optimizations), 
properties and system state as conceptually
shown in Figure~\ref{fig:universal-node}. At a coarse-grain level, modules serve 
as wrappers around existing command line tools such as compilers, source-to-source 
transformers, code launchers, profilers, among many others. Such modules are written 
in python for productivity and portability reasons, and can be launched from command line 
in a unified way using Collective Mind front-end \emph{cm} as following:

{
\center
\small
\emph{
cm $\langle$ module name or UID $\rangle$  $\langle$ command $\rangle$  
$\langle$ unified meta information $\rangle$  
-- 
$\langle$ original cmd $\rangle$ 
}
}

These modules enable transparent monitoring of information flow, exposure
of various characteristics and properties in a unified way (meta information),
and exploration or prediction of design and optimization choices, while helping
researchers to abstract their experimental setups from constant changes in
the system. Internally, modules can call each other using just one unified \emph{cM access function} 
which uses a schema-free easily extensible nested dictionary that can be directly serialized 
to JSON as both input and output as following:

{
\begin{small}
\begin{verbatim}
r=cm_kernel.access({'cm_run_module_uoa':<module name or UID>,
                      'cm_action':<command>,
                      parameters})
if r['cm_return']>0: 
   print 'Error:'+r['cm_error']
   exit(r['cm_return'])
\end{verbatim}
\end{small}
}
where command in each module is directly associated with some function.
Since JSON can also be easily transmitted through Web using standard http post mechanisms,
we implemented a simple cM web server that can be used for P2P communication
or centralized repository during crowdsourcing and possibly multi-agent based 
on-line learning and tuning.

Each module has an associated storage 
that can preserve any collections of files 
(whole benchmark, data set, tool, trace, model, etc) and their 
meta-description in a JSON file. Thus each module can also be used
for any data abstraction and includes various common commands standard to any 
repository such as \emph{load, save, list, search, etc}. We use our own simple directory-based format as following:
{
\begin{small}
\begin{verbatim}
.cmr/<Module name or UID>/<Data entry UID>
\end{verbatim}
\end{small}
} where .cmr is an acronym for Collective Mind Repository.
In contrast with using SQL-based database in the first cTuning version that
was fast but very complex for data sharing or extensions of structure and relations,
a new open format allows users to be database and
technology-independent with the possibility to modify, add, delete or share
entries and whole repositories using standard OS functions and tools like
SVN, GIT or Mercury, or easily convert them to any other format or database
when necessary. Furthermore, cM can transparently use open source
JSON-based indexing tools such as ElasticSearch~\cite{software-elastic-search}
to enable fast and powerful queries over schema-free meta information.
Now, any research artifact will not be lost and can now be referenced and directly found 
using the so called cID (Collective ID) of the format: \emph{$\langle$ module name 
or UID $\rangle$:$\langle$ data entry or UID $\rangle$}.

Such infrastructure allows researchers and engineers to connect existing 
or new modules into  experimental pipelines like "research LEGO" 
with exposed characteristics, properties, constraints and states to 
quickly and collaboratively prototype and crowdsource their ideas 
or production scenarios such as traditional adaptive exploration 
of large experimental spaces, multi-objective program and architecture 
optimization or continuous on-line learning and run-time adaptation
while easily utilizing all available benchmarks, data sets, tools and models
provided by the community. Additionally, single and unified access function
enables transparent reproducibility and validation of any experiment 
by preserving input and output dictionaries for a given experimental 
pipeline module. Furthermore, we decided to keep all modules inside repository
thus substituting various ad-hoc scripts and tools. With an additional 
cM possibility to install various packages and their dependencies automatically (compilers,
libraries, profilers, etc) from the repository or keep all produced binaries
in the repository, researchers now have an opportunity to preserve and share the 
whole experimental setup in a private
or public repository possibly with a publication.

\begin{figure}[ht]
  \centering
  \includegraphics[width=3in, angle=-90, bb=64 301 525 540]{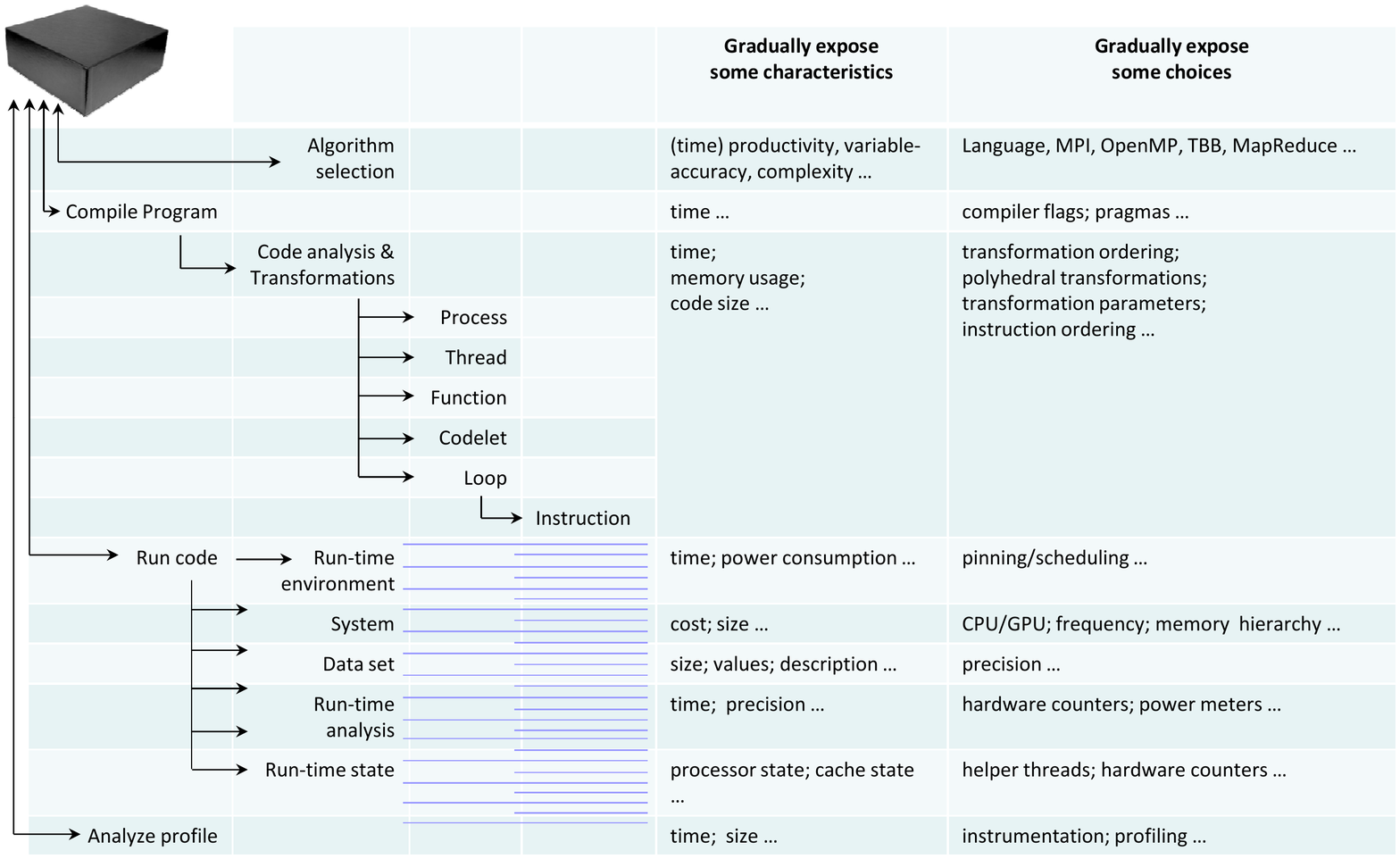}
  \caption{\it Gradual top-down decomposition of computer systems to balance coarse-grain vs. fine-grain analysis and tuning depending on user requirements and expected ROI}
  \label{fig:top-down-analysis}
\end{figure}

We started collaborative and gradual decomposition of large, coarse-grain components 
into simpler sub-modules including decomposition of programs into kernels 
or codelets~\cite{Zuckerman:2011:UCP:2000417.2000424} as shown in Figure~\ref{fig:top-down-analysis}
to keep complexity under control and possibly use multi-agent based or brain 
inspired modeling and adaptation of the behavior of the whole computer system locally or during
P2P crowdsourcing. Such decomposition also allows community to first learn and optimize
coarse-grain behavior, and later add more fine-grain effects depending on
user requirements, time constraints and expected return on investment
(ROI) similar to existing analysis methodologies in physics, electronics or finances.
	
\subsection{Data and parameter description and classification}

In traditional software engineering, all software components and their
API are usually defined at the beginning of the project to avoid modifications later. 
However, in our case, due to ever evolving tools, APIs and data formats, we decided 
to use agile methodology together with type-free inputs and outputs for 
all functions focusing on quick and simple prototyping of research ideas. 
Only when modules and their inputs and outputs become mature or validated, 
then (meta)data and interface are defined, systematized and classified. 
However, they can still be extended and reclassified at any time later.

For example, any key in an input or output dictionary of a given function
and a given module can be described as "choice", "(statistical) characteristic", 
"property" and "state", besides a few internal types including "module UID"
or "data UID" or "class UID" to provide direct or semantic class-based 
connections between data and modules. Parameters can be discrete or continuous with a given range to enable
automatic exploration. Thus, we can easily describe compiler optimizations;
dataset properties such as image or matrix size, architecture properties
such as cache size or frequency, represent execution time, 
power consumption, code size, hardware counters; 
categorize benchmarks and codelets in terms of reaction to optimizations
or as CPU or memory bound, and so on. 


\subsection{OpenME interface for fine-grain analysis, tuning and adaptation}

\begin{figure}[ht]
  \centering
  \includegraphics[width=5in, bb=73 301 636 540]{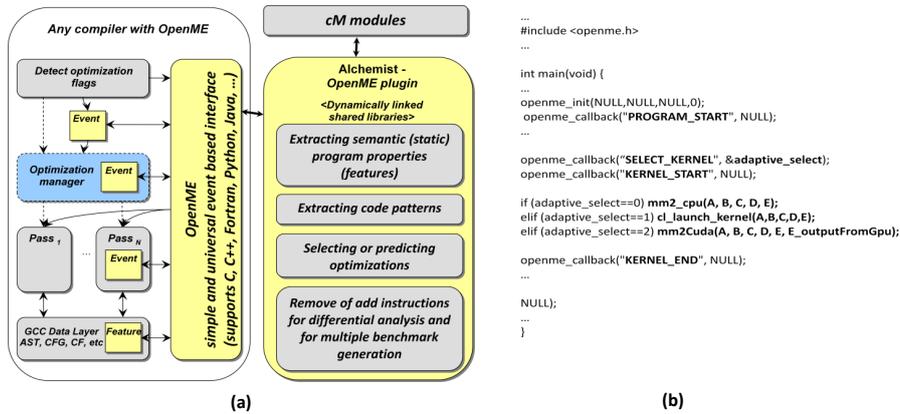}
  \caption{\it Event and plugin-based OpenME interface to "open up" rigid tools (a) and applications (b)
for external fine-grain analysis, tuning and adaptation, and connect them to cM}
  \label{fig:openme}
\end{figure}

Most of current compilers, applications and run-time systems are not 
prepared for easy and straightforward fine-grain analysis and tuning 
due to associated software engineering complexity, sometimes 
proprietary internals, possible compile or run-time overheads,  
and still occasional disbeliefs in effective run-time adaptation. 
Some extremes included either fixing, hardwiring and hiding 
all optimization heuristics from end-users or oppositely 
exposing all possible optimizations, scheduling parameters, 
hardware counters, etc. Some other available mechanisms 
to control fine-grain compiler optimization through pragmas 
can also be very misleading since it is not always easy or possible 
to validate whether optimization was actually performed or not.

Instead of developing yet another source-to-source tools or binary translators and
analyzers that always require enormous resources and effort often to reimplement 
functionality of existing and evolving compilers and support evolving architectures, 
we developed a simple event and plugin-based interface called Interactive Compilation
Interface (ICI) to "open up" previously hardwired tools for external analysis
and tuning. ICI was written in plain C originally for Open64 and 
later for GCC, requires minimal instrumentation of a compiler
and helps to expose or modify only a subset of program properties or compiler
optimization decisions through external dynamic plugins based on 
researcher needs and usage scenario. This interface can easily evolve with the compiler
itself, has been successfully used in the MILEPOST project to build
machine-learning self-tuning compiler~\cite{29db2248aba45e59:a31e374796869125}, 
and is now available in mainline GCC. 

Based on this experience, we developed a new version of this interface (OpenME)~\cite{software-cm}
that is used to "open up" any available tool such as
GCC, LLVM, Open64, architecture simulator, etc in a unified way as shown in
Figure~\ref{fig:openme}(a), or any application for example to train
predictive scheduler on heterogeneous many-core
architectures~\cite{JGVP2009} as shown in Figure~\ref{fig:openme}(b). It
can be connected to cM to monitor application behavior in realistic
environments or utilize on-line learning modules to quickly prototype
research ideas when developing self-tuning applications that can
automatically adapt to different datasets, underlying architectures
particularly in virtual and cloud environments, or react to changes in
environment and run-time behavior. Since there are some natural overheads associated with event
invocation, users can substitute them with hardwired fast calls after
research idea has been validated. We are developing associated Alchemist
plugin~\cite{software-cm} for GCC to extract code structure, patterns and various
properties to substitute and unify outdated MILEPOST GCC plugin~\cite{milepost_gcc_webpage} for
machine-learning based meta compilers.

\section{Possible usage scenarios}

We decided first to re-implement various analysis, tuning and learning scenarios 
from our past research as cM modules combined into universal compilation and execution 
pipeline to give the community a common reproducible base for further research and 
experimentation. Furthermore, rather than just showing speedups, our main focus is also to 
use our distributed framework similar to web crawlers to search for unusual or
unexpected behavior that can be exposed to the community for further analysis, advise,
ranking, commenting, and eventual improvement of cM to take such behavior into account.

\subsection{Collaborative observation and exploration}

\begin{figure}[ht]
  \centering
  \includegraphics[width=4.4in, bb=73 348 686 540]{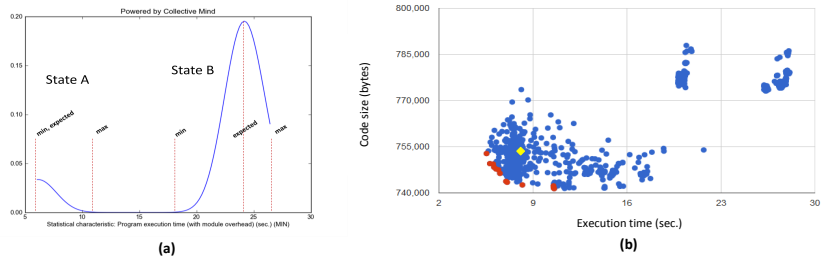}
  \caption{\it (a) execution time variation of a susan corner codelet with the same dataset on Samsung Galaxy Y for 2 frequency states; (b) variation in execution time vs code size during GCC 4.7.2 compiler flag auto-tuning for the same codelet on the same mobile phone where yellow rhombus represents -O3 and red circles show Pareto frontier - all data and modules are available for reproduction at c-mind.org/repo}
  \label{fig:analysis}
\end{figure}

Having common and portable framework with exposed characteristics, properties, choices
and state in a unified way allows us to collaboratively observe,
optimize, learn and predict program behavior on any existing system including
mobile phones, tablets, desktops, servers, cluster or cloud nodes
in realistic environment transparently and on the fly instead of using a
few ad-hoc and often non-representative benchmarks, data sets and platforms.
Furthermore, it elegantly solves a common problem of a lack of experimental data 
to be able to properly apply machine learning techniques and make
statistically meaningful assumptions that slowed down many recent projects on
applying machine learning to compilation and architecture.

For example, Figure~\ref{fig:analysis}(a) shows variation of an execution time
of an image corner detection codelet with the same dataset (image) on a Samsung Galaxy Y 
mobile phone with ARM processor using modified Collective Mind Node~\cite{software-cmn}.
Our cM R-based statistical module reports that distribution is not normal that
usually results in discarding this experiment in most of the research projects.
However, by exposing and analyzing this relatively simple case, we found that processor frequency
was responsible for this behavior thus adding it as a new parameter to the "state" vector of our
experimental pipeline to effectively separate such cases. Furthermore, we
can use minimal execution time for SW/HW co-design as the best what a
given code can achieve on a given architecture, or expected execution time for
realistic end-user program optimization and adaptation.

\subsection{Adaptive exploration (sampling) and dimensionality reduction}

Now, we can easily distribute exploration of any set of choices vs multiple properties and
characteristics in a computer system among many users. For example, Figure~\ref{fig:analysis}(b) shows 
random exploration of Sourcery GCC compiler flags versus execution time and binary size 
on off-the-shelf Samsung mobile phones with ARMv7 processor for image processing
codelet while Figure~\ref{fig:universal-learning} shows exploration of dataset parameters 
for LU-decomposition numerical kernel on GRID5000 machines with Intel Core2 and SandyBridge processors.
Since all characteristics are usually dependent, we can apply cM plugin (module)
to detect universal Paretto fronter on the fly in multi-dimensional space (currently not optimal) 
during on-line exploration and filter all other cases. A user can choose to explore any other available characteristic
in a similar way such as power consumption, compilation time, etc depending on usage
scenario and requirements. In order to speed up random exploration further, we use
probabilistic focused search similar to ANOVA and PCA 
described in~\cite{FOTP2005,29db2248aba45e59:530e5f456ea259de} that can suggest most important tuning/analysis 
dimensions with likely highest speedup or unusual behavior, and guide further finer-grain 
exploration in those areas. Collective exploration is critical to build and update 
a realistic training set for machine-learning based self-tuning meta-compiler cTuning-CC
to automatically and continuously improve default optimization heuristic of GCC, LLVM, 
ICC, Open64 or any other compiler connected to cM~\cite{29db2248aba45e59:a31e374796869125,milepost}.

\subsection{On-line learning}

\begin{figure}[ht]
  \centering
  \includegraphics[width=4.4in, bb=0 0 648 360]{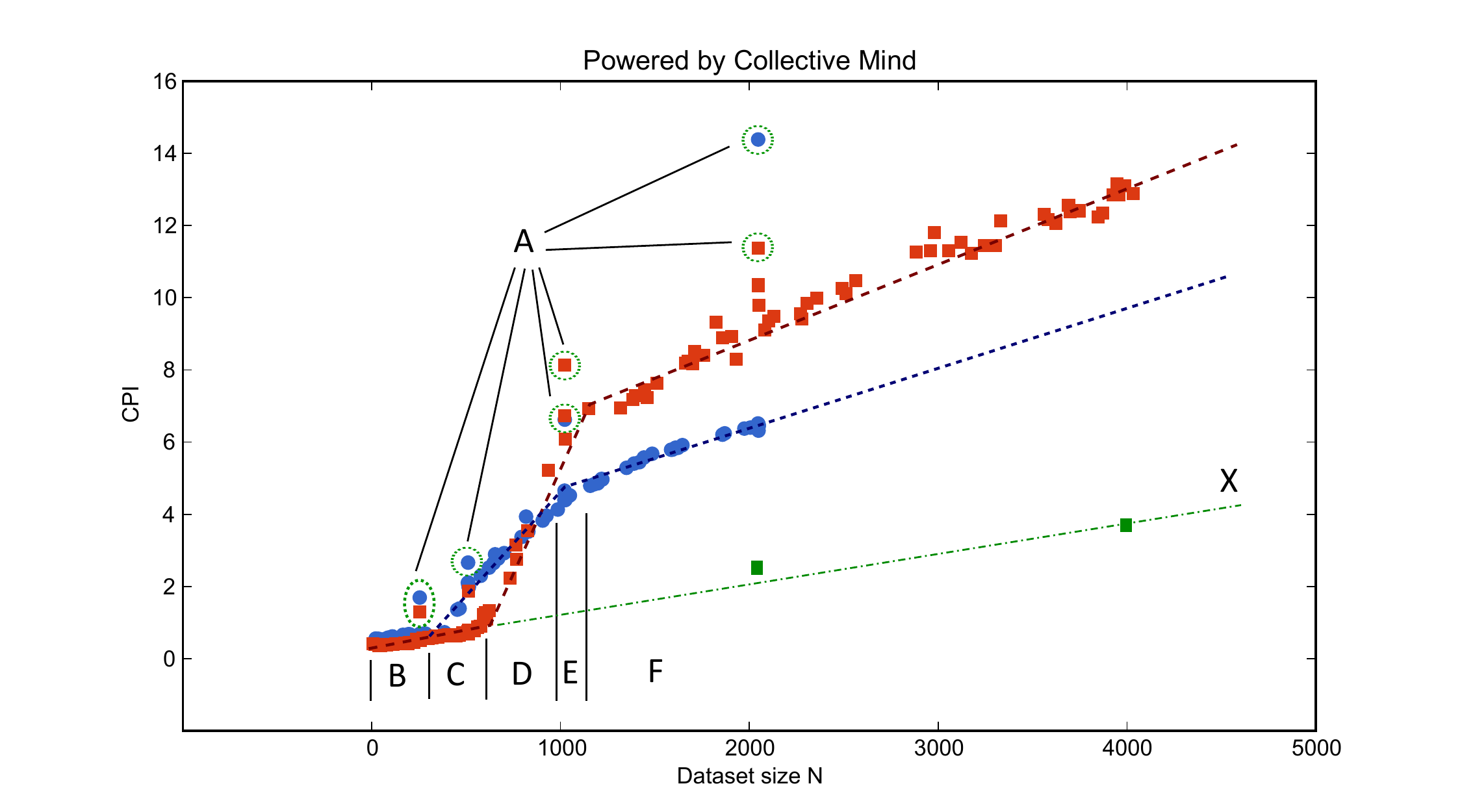}
  \caption{\it On-line learning (predictive modeling) of a CPI behavior of ludcmp on 2 different platforms (Intel Core2 vs Intel i5) vs matrix size N and cache size}
  \label{fig:universal-learning}
\end{figure}

Crowdtuning has a side effect - generation and processing of huge amount of data 
that is well-known in other fields as a "big data" problem. However,
in our past work on online tuning, we showed that it is possible not only to 
learn behavior and find correlations between characteristics, properties
and choices to build models of behavior on the fly at each client or program, 
but also to effectively compact experimental data keeping only representative 
or unexpected points, and minimize communications between cM nodes thus making cM a giant, 
distributed learning network to some extent similar to brain~\cite{FCOT2005,LCWP2009,29db2248aba45e59:530e5f456ea259de}.

Figure~\ref{fig:universal-learning} demonstrates how on-line learning is
performed in our framework using LU-decomposition benchmark as an example,
CPI characteristic, and 2 Intel-based platforms (Intel Core2
Centrino T7500 Merom 2.2GHz L1=32KB 8-way set-associative, L2=4MB 16-way
set associative - red dots vs. Intel Core i5 2540M 2.6GHz Sandy Bridge
L1=32KB 8-way set associative, L2=256KB 8-way set associative, L3=3MB
12-way set associative - blue dots). At the beginning, our system does 
not have any knowledge about behavior of this (or any other) benchmark, 
so it simply observes and stores available characteristics while collecting 
as many properties of the whole system as possible (exposed by a researcher or user). 
At each step, system processes all historical observations using various 
available predictive models such as SVM or MARS in order to find correlations
between properties and characteristics. In our example, after sufficient amount 
of observations, system can build a model that automatically correlated 
data set size N, cache size and CPI (in our case combination of linear models 
B-F that reflect memory hierarchy of a particular system) with nearly 100\%
prediction. However, system always continue observing behavior to 
continuously validate it against existing model in order to detect discrepancies
(failed predictions). In our case, the system eventually detects outliers A that are
due to cache alignment problems. Since off-the-shelf models rarely handle
such cases, our framework allows to exclude such cases from modeling and
expose them to the community through the unified Web services to reproduce 
and explain this behavior, find relevant features and improve or optimize 
existing models. In our case, we managed to fit a hybrid rule-based model 
that first validates cases where data set size is a power of 2, otherwise 
it uses linear models as functions of a data set and cache size.


Systematic characterization (modeling) of a program behavior across many systems
and data sets allows researchers or end-users to focus further optimizations (tuning) 
only on representative areas with distinct behavior while collaboratively building 
an on-line advice system. In the above example, we evaluated and prepared the
following advices for optimizations: points A can be optimized using array
padding; area B can profit from parallelization and traditional compiler
optimizations targeting ILP; areas C-E can benefit from loop tiling; area F
and points A can benefit from reduced processor frequency to reduce
power consumption using cM online adaptation plugin. Since auto-tuning is continuously 
performed, we will  release final optimization details at cM live repository~\cite{cm-repo} 
during symposium.             

\begin{figure}[ht]
  \centering
  \includegraphics[width=4.4in, bb=14 8 716 524]{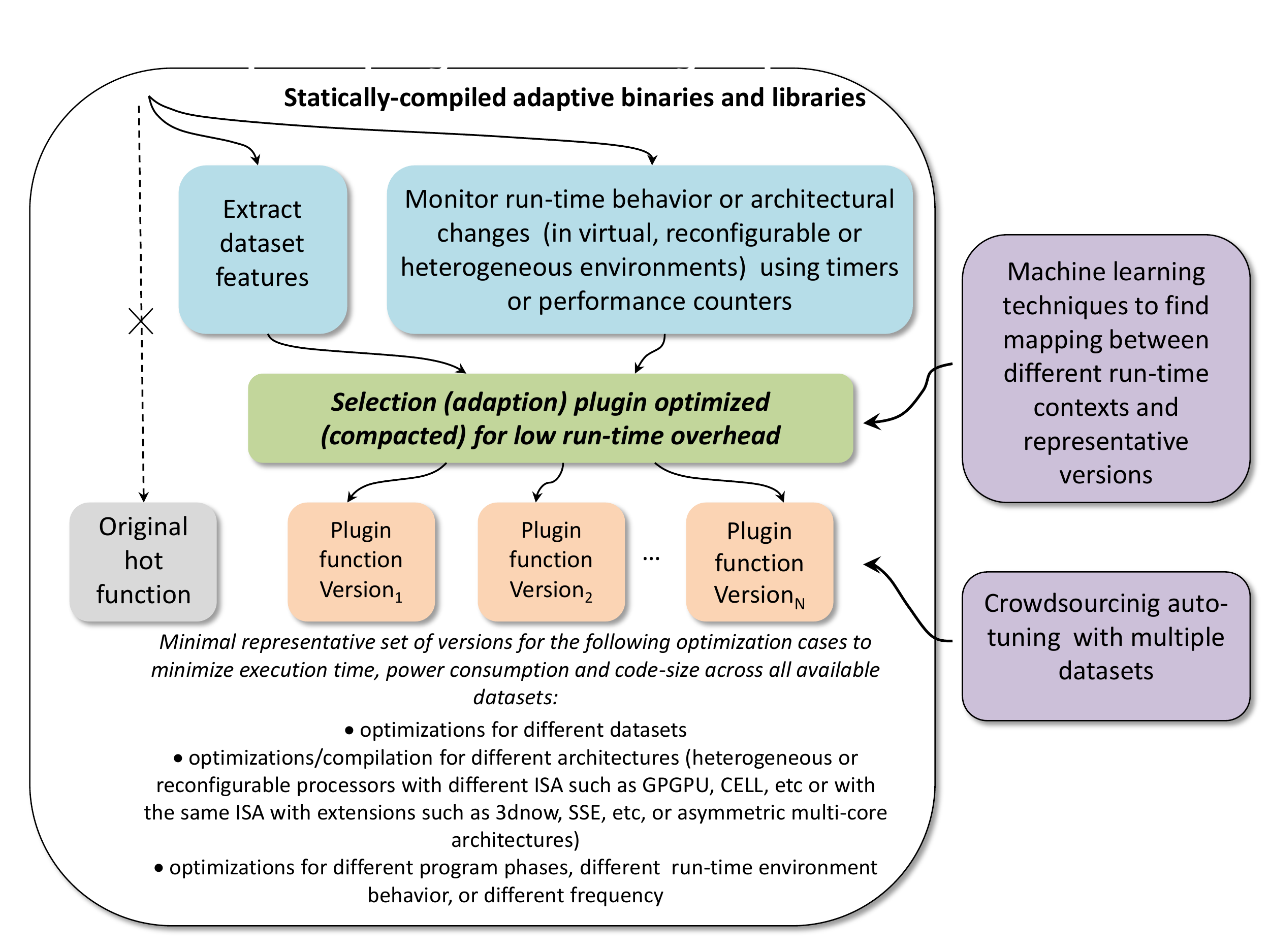}
  \caption{\it Systematizing and unifying split (staged) compilation for statically built adaptive applications using crowdtuning and machine learning}
  \label{fig:multi-versions}
\end{figure}

Gradually increasing and systematized knowledge in the repository in form of models 
can now be used to detect and characterize an abnormal program or system behavior, 
suggest future architectural improvements, or predict most profitable program 
optimizations, run-time adaptation scenarios and architecture configurations 
depending on user requirements. For example, this knowledge can be effecitvely 
used for split (staged) tuning to build static multi-versioning applications
with cM plugins for phase-based adaptation~\cite{FCOT2005}  or predictive scheduling~\cite{JGVP2009} 
in heterogeneous systems that can automatically adjust their behavior at run-time to varying data sets, environments, architectures and system state by selecting 
appropriate versions or changing frequency to maximize performance and minimize power consumption,
while avoiding complex recompilation frameworks as conceptually shown in Figure~\ref{fig:multi-versions}.

\subsection{Benchmark automatic generation and decremental analysis}

Projects on applying machine learning to auto-tuning suffer from yet another
well-known problem: lack of benchmarks. Our experience with hundreds of codelets 
and thousands of data sets~\cite{FCOP2007,Chen10evaluatingiterative,29db2248aba45e59:a31e374796869125} 
shows that they are still not enough to cover all possible properties and behavior 
of computer systems. Generating numerous synthetic benchmarks and data sets 
is theoretically possible but will result in additional explosion in analysis and tuning dimensions. 
Instead, we use existing benchmarks, codelets and even data sets as templates,
and utilize Alchemist plugin~\cite{software-cm} for GCC to randomly or incrementally modify 
them by removing, modifying or adding various instructions, basic blocks, loops, 
and thus generating. Naturally, we ignore crashing variants of the code and continue 
evolving only the working ones. Importantly, we use this approach not only to extend
realistic training sets, but also to gradually identify various behavior anomalies 
and detect code properties to explain these anomalies and improve predicting modeling
without any need for slow and possibly imprecise system/architecture simulator 
or numerous and sometimes misleading hardware counters as originally presented in~\cite{FOTP2004,29db2248aba45e59:61df8131994fad97}. 
For example, we can scalarize memory accesses to characterize code and data set
as CPU or memory bound~\cite{29db2248aba45e59:61df8131994fad97} (line X in Figure~\ref{fig:universal-learning} 
shows ideal codelet behavior when all floating point memory accesses are NOPed). Additionally, we use Alchemist plugin to extract 
code structure, patterns and other properties to improve our cTuning CC machine-learning 
based meta compiler connected to GCC, LLVM, Open64, Intel and Microsoft compilers,
and to guide SW/HW co-design.




\section{Conclusions and future work}

With the continuously rising number of workshops, conferences,
journals, symposiums, consortiums, networks of excellence, publications,
tools and experimental data, and at the same time decreasing number of
fundamentally new ideas and reproducible research in computer engineering,
we strongly believe that the only way forward now is to start collaborative
systematization and unification of available knowledge about design and
optimization of computer systems. However, unlike some existing projects 
that mainly suggest or attempt to share raw experimental data and related tools, 
and somehow validate results by the community, or redesign the whole software
and hardware stack from scratch, we use our interdisciplinary background 
and experience to develop the first to our knowledge integrated, extensible
and collaborative infrastructure and repository (Collective Mind) that can 
represent, preserve and connect directly or semantically all research artifacts 
including data, executable code and interfaces in a unified way. 

We hope that our collaborative, evolutionary and agile methodology, and extensible 
plugin-based Lego-like framework can help to address current fundamental challenges in computer
engineering while bringing together interdisciplinary communities similar to Wikipedia
to continuously validate, systematize and improve collective knowledge about
designing and optimizing whole computer systems, and extrapolate it to
build faster, more power efficient, reliable and adaptive devices and
software. We hope that community will continue developing more plugins (modules) 
to plug various third-party tools including TAU~\cite{Shende:2006:TPP:1125980.1125982}, Periscope~\cite{periscope}, 
Scalasca~\cite{Geimer:2010:SPT:1753228.1753234}, Intel vTune
and many others to cM, or continue gradual decomposition of programs into codelets and complex tools
into simpler connected self-tuning modules while crowdsourcing learning, tuning and classifying  of their
behavior. We started building a large public repository of realistic behavior 
of multiple programs in realistic environments with
realistic data sets ("big data") that should allow the community to quickly
reproduce and validate existing results, and focus their effort on
developing novel tuning techniques combined with data mining,
classification and predictive modeling rather than wasting time on
building individual experimental setups. It can also be used to address the
challenge of collaboratively finding minimal representative set of benchmarks, codelets and datasets 
covering behavior of most of existing computer systems, detecting correlations in a collected data
together with combinations of relevant properties (features), pruning
irrelevant ones, systematizing and compacting existing experimental data,
removing or exposing noisy or wrong experimental results. It can also be effectively used to validate and compact existing
models including roofline~\cite{Williams:2009:RIV:1498765.1498785} or capacity~\cite{Kuck2012} ones, and adaptation techniques including multi-agent based using cM P2P communication,
classify programs by similarities in models, by reactions to optimizations~\cite{29db2248aba45e59:530e5f456ea259de} and to semantically non-equivalent changes~\cite{FOTP04}, 
or collaboratively develop and optimize new complex hybrid predictive models that from our past practical experience
can not yet be fully automated thus using data mining and machine learning 
as a helper rather than panacea at least at this stage.

Beta proof-of-concept version of a presented infrastructure and its documentation 
is available for download at~\cite{software-cm}, while pilot
Collective Mind repository is now live at \emph{c-mind.org/repo}~\cite{cm-repo}
and currently being populated with our past research artifacts including
hundreds of codelets and benchmarks~\cite{29db2248aba45e59:a31e374796869125}, thousands of data
sets~\cite{FCOP2007,Chen10evaluatingiterative}, universal compilation and execution pipeline with adaptive
exploration (tuning), dimension reduction and statistical analysis modules,
and classical off-the-shelf or hybrid predictive models. 
Importantly, presented concepts have already been successfully validated in several academic and industrial projects with
IBM, ARC (Synopsys), CAPS, CEA and Intel, and we gradually release all our
experimental data from these projects including unexplained behavior of
computer systems and misbehaving models. Finally, the example of a Collective Mind Node to crowdsource
auto-tuning and learning using Android mobile phones and tables is available at Google Play~\cite{software-cmn}.

We hope that our approach will help to shift current focus from publishing only good
experimental results or speedups, to sharing all research artifacts,
validating past techniques, and exposing  unexplained behavior or
encountered problems to the interdisciplinary community for reproducibility and collaborative solving
and ranking. We also hope that Collective Mind framework will be of help to a broad range of researchers
even outside of computer engineering not to drawn in their experimentation while processing, systematizing, and sharing
their scientific data, code and models. Finally, we hope that Collective Mind
methodology will help to restore the attractiveness of computer engineering
making it a more systematic and rigorous discipline~\cite{fur2013_hipeac_ts_mcs}. 

\section*{Acknowledgments}

{\small
Grigori Fursin was funded by EU HiPEAC postdoctoral fellowship (2005-2006) 
and by the EU MILEPOST project (2007-2010) where he originally developed and applied his statistical 
plugin-based crowdtuning technology to enable realistic and on-line tuning (training) 
of a machine-learning based meta-compiler cTuning CC together with MILEPOST GCC,
and later by French Intel/CEA Exascale Lab (2010-2011) where he extended
this concept and developed customized codelet repository and auto-tuning 
infrastructure for software and hardware co-design and co-optimization of Exascale systems 
together with his team (Yuriy Kashnikov, Franck Talbart, and Pablo Oliveira). 
Grigori is grateful to Francois Bodin and CAPS Entreprise for sharing codelets
from the MILEPOST project and for providing an access to the latest Codelet
Finder tool, to David Kuck and David Wong from Intel Illinois, and Davide del Vento 
from NCAR for interesting discussions and feedback during development of cTuning technology. 
Grigori is also very thankful to cTuning and HiPEAC communities as well as his colleagues from
ARM and STMicroelectronics for motivation, many interesting discussions and
feedback during development of Collective Mind repository and
infrastructure presented in this paper. Finally he would like to thank
GRID5000 project and community for providing an access to powerful
computational resources.
}

\bibliographystyle{abbrv}
\bibliography{paper}
\end{document}